\documentclass[aps,pra,twocolumn,showpacs,superscriptaddress,groupedaddress]{revtex4}
\usepackage{graphicx} 
\usepackage{dcolumn}   
\usepackage{bm}    
\usepackage{amssymb}   
\usepackage{epstopdf}
\usepackage{amsmath}
\begin{document}
\preprint{APS/123-QED}
\title{Entanglement fidelity for electron-electron interaction in strongly coupled semiclassical plasma and under  external fields}
\author{B. J. Falaye}
\email{babatunde.falaye@gmail.com}
\affiliation{Applied Theoretical Physics Division​, Department of Physics, Federal University Lafia,  P. M. B. 146, Lafia, Nigeria.}
\author{K. J. Oyewumi}
\affiliation{Theoretical Physics Section, Department of Physics, University of Ilorin,  P. M. B. 1515, Ilorin, Nigeria.}
\author{O. A. Falaiye}
\affiliation{Theoretical Physics Section, Department of Physics, University of Ilorin,  P. M. B. 1515, Ilorin, Nigeria.}
\author{C. A. Onate}
\affiliation{Physics Programme, Department of Physical Sciences, Landmark Univerisity, Omu-Aran, Nigeria}
\author{O. J. Oluwadare}
\affiliation{Department of Physics, Federal University Oye-Ekiti, P. M. B. 373, Ekiti State, Nigeria}
\author{W. A. Yahya}
\affiliation{Department of Physics and Material Science, Kwara State University, P. M. B. 1530, Ilorin, Nigeria}

\begin{abstract}
\noindent
This paper presents the effects of AB-flux field and electric field on electron-electron interaction, encircled by a strongly coupled semiclassical plasma. We found that weak  external fields are required to perpetuate a low-energy elastic electron-electron interaction in a strongly coupled semiclassical plasma. The entanglement fidelity in the interaction process has been examined. We have used partial wave analysis to derive the entanglement fidelity. We found that for a weak electric field, the fidelity ratio for electron-electron interaction increase as projectile energy increase but remains constant or almost zero for a strong electric field. Our results provide an invaluable information on how the efficiency of entanglement fidelity for a low-energy elastic electron-electron interaction in a strongly coupled semiclassical plasma can be influenced by the presence of external fields. 
\end{abstract}
\maketitle

\section{Introduction}
\label{sec1}
There has been a ceaseless interest (\cite{A1,A2,A3} and Refs. therein) in studying electron-ion and electron-electron interaction process in plasmas environment due to its application in diagnosing various plasma and also providing passable knowledge on collision dynamics. It was shown in Ref. \cite{A3} that the optically allowed but 1s-2s cross section is less sensitive to the plasma effect but 1s-2p and 2s-2p cross sections for electron impact excitation are substantially reduced by the plasma screening. The effect of the carrier reservoir dimensionality on electron-electron scattering in quantum dot materials was investigated in Ref. \cite{ER1}.  It was found that 3D scattering benefits from its additional degree of freedom in the momentum space. Jung \cite{ER2} studied quantum-mechanical effects on electron-electron scattering in dense high-temperature plasmas. Suppression of inelastic electron-electron scattering in Anderson insulators was explored by Ovadyahu \cite{ER3}.

The study of entanglement has been a subject of active research in physics, chemistry and related areas. Quantum entanglement lies in the heart of quantum mechanics (\cite{TE1,TE2,TE3} and refs. therein). Entanglement has become very useful in information processing and quantum communication. Really, it plays a crucial role in potential realization of quantum computers. It also represent a quantitative measure for electron-electron  correlation in many body system. A very useful tool to distinguish between two quantum states (say $\rho_1$ and $\rho_2$) is known as the fidelity between $\rho_1$ and $\rho_2$, i.e., $F(\rho_1,\rho_2):=$Max$|\left\langle \phi_1\right.|\left.\phi_2\right\rangle|$, where  $\left.|\phi_1\right\rangle$ and $\left.|\phi_2\right\rangle$ correspond to the state of $\rho_1$ and $\rho_2$.

The interest in studying entanglement fidelity in the electron-electron scattering process  has received a considerable attention in the last few years. This is due to the key role that quantum correlation plays in understanding information processing. The entanglement in scattering processes was investigated by Mishima et al. \cite{JA11}. It has been found that, a low energy elastic scattering is desirable for entanglement enhancement. The quantum effect on the entanglement fidelity in elastic electron-ion scattering was investigated under strongly coupled semiclassical plasmas in Ref. \cite{A7}. It was shown that the quantum effect significantly augments the entanglement fidelity in a strongly coupled semiclassical plasmas. In the present work, we study the conditions to obtain a low-energy elastic electron-electron scattering by taking into account the effect of electric field, AB-flux field and  uniform magnetic field directed along $z$-axis and surrounded by strongly coupled semiclassical plasmas. Once the conditions to obtain a low energy level are established, then entanglement fidelity will be examined.

\section{Theory and Calculations}
\label{sec2}
The Debye-H\"uckel model which provides a modern treatment of non-ideality in plasma via the screening effect is given by $\chi(r)=Z_aZ_b/r\exp(-r/\lambda_D)$,  where $\lambda_D$ represent Debye length given by $\lambda_D = \sqrt{k_\beta T/4\pi e^2\sum_in_iZ_i^2}$ with $Z_a$ and $Z_b$ being the $a-b$ interaction (here, $a=b=e$) . This model accounts for pair correlations. However, besides correlation, the quantum mechanical effects of diffraction also take place in semi classical plasma. Hence, the effective model which accounts for both screening and diffraction in strongly coupled semiclassical plasmas can be written as \cite{A9}: 
\begin{eqnarray}
\chi(r)&=&-\frac{Z_aZ_be^2}{\sqrt{1-4\frac{\lambda_{ab}^2}{\lambda_D^2}}}\left[\mathcal{B}^2(\lambda_{ab},\lambda_D)\frac{\exp[-\mathcal{A}(\lambda_{ab},\lambda_D)r]}{r}\right.\nonumber\\
&&\left.-\mathcal{A}^2(\lambda_{ab},\lambda_D)\frac{\exp[-\mathcal{B}(\lambda_{ab},\lambda_D)r]}{r}\right], 
\end{eqnarray}
where $\mathcal{A}$ and $\mathcal{B}$ are quantum screening parameters given by $\mathcal{A}^2\equiv\left(1-\sqrt{1-4\lambda^2_{ab}/\lambda_D^2}\right)/2$, $\mathcal{B}^2\equiv\left(1+\sqrt{1-4\lambda_{ab}^2/\lambda_D^2}\right)/2$ and the de Broglie thermal wavelength is denoted by $\lambda_{ab}$. 

On this foundation, we can build the model equation for electron-electron interaction under the influence of electric field, AB-flux field and uniform magnetic field directed along $z$-axis and surrounded by strongly coupled semiclassical plasmas, in spherical polar coordinates as:
\begin{widetext}
\begin{equation}
\left[\frac{-\hbar^2}{2\mu}{\nabla}^2-\underbrace{\frac{\left(Z_ee\right)^2}{\sqrt{1-4\frac{\lambda_{ee}^2}{\lambda_D^2}}}\left[\mathcal{B}^2\frac{e^{-\mathcal{A}(\lambda_{ee},\lambda_D)r}}{r}-\mathcal{A}^2\frac{e^{-\mathcal{B}(\lambda_{ee},\lambda_D)r}}{r}\right]}_{\mbox{screened effective
pseudopotential model}}-Fr\cos(\theta)\right]\psi(r,\theta)=E_{nm}\psi(r,\theta),
\label{EE1}
\end{equation}
\end{widetext}
where $E$ denotes the eigenvalues, $\mu$ is the effective mass of an electron,  vector potential $\vec{A}=\frac{\phi_{AB}}{2\pi r}\hat{\phi}$ represents the magnetic flux $\phi_{AB}$ created by a solenoid inserted inside the antidot with  $\vec{\nabla}\times\vec{A}=0$. $\lambda_D$ is the screening parameter. $Z$ denotes the atomic number which is found useful in describing energy levels of light to heavy neutral atoms. The thermal de Broglie wavelength of the electron-electron pair is denoted by $\lambda_{ee}=\hbar/\sqrt{\pi\mu k_\beta T_e}$ where $T_e$ is the temperature of the plasma electrons, Planck constant is denoted as $\hbar$ and $k_\beta$ is the Boltzmann constant. Moreover, the characteristic properties of the plasma are denoted  by the coupling parameter $\alpha k_\beta T_e\Gamma_{ee}=(Ze)^2$ (where $\alpha$ is the average distance between particles). $\lambda_D=\sqrt{K_\beta T_e/4\pi n_ee^2}$ is the Debye radius where the electron density is denoted by $n_e$. The ranges of electron density and temperature, $T_e$ are known as $~10^{20}-10^{24}$cm$^{-3}$ and $~5\times10^4-10^6K$, respectively in dense classical plasma.

\begin{figure*}[!t]
\includegraphics[width=\linewidth]{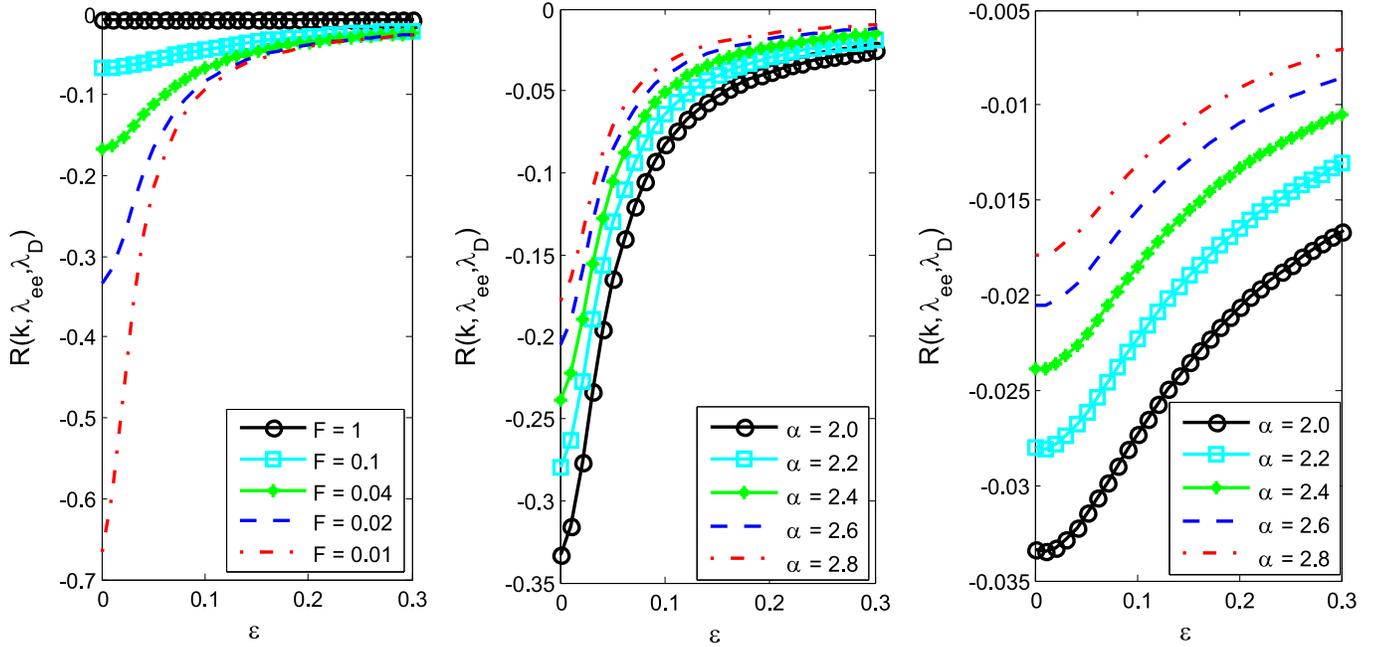}
\caption{{\protect\footnotesize Variation of fidelity ratio for electron-electron interaction under the influence of electric field in atomic units using the fitting parameters $\alpha=2=r_s=2$ $\Gamma_{ee}=1$, $\delta=5$, $\Gamma_{ee}=1$ (a) as a function of the scaled projectile energy for various values of F. (b) as a function of the scaled projectile energy for various values of the density parameter and with $F=0.02$(c) Same as (b) but for $F=0.2$.}}
\label{fig1}
\end{figure*}

Furthermore, $F$ represents electric field strength with angle $\theta$ between $F$ and $r$. With $\theta=0$, then $F r\cos(\theta)$ becomes $Fr$. The variation of the effective potential energy as a function of various model parameters have been displayed in figure  \ref{fig1}. Now, let us take a wavefunction in cylindrical coordinates as $\psi(r,\phi)=\frac{1}{\sqrt{2r\pi}}e^{im\phi}\mathcal{H}_{nm}(r)$, where $m=0, \pm1, \pm2,...$ denotes the magnetic quantum number. Inserting this wavefunction into equation (\ref{EE1}), we find a second order differential equation ${d^2\mathcal{H}_{nm}(r)}/{dr^2}+{2\mu}/{\hbar^2}\left[E_{nm}-\chi_{\rm eff.}\right]\mathcal{H}_{nm}(r)=0$, where the effective potential V$_{\rm eff.}$ is
\begin{widetext}
\begin{eqnarray}
\chi_{\rm eff.}&=&-\frac{\alpha\Gamma_{ee}}{\sqrt{1-\frac{24\Gamma_{ee}^2}{\pi r_s\left(1+\delta^2\right)}}}\left[A_{-}^2\frac{e^{-A_{+}r}}{r}-A_{+}^2\frac{e^{-A_{-}r}}{r}\right]-Fr+\frac{\hbar^2}{2\mu}\left[\frac{\left(\sigma_{0m}-\frac{1}{2}\right)^2-\frac{1}{4}}{r^2}\right]\nonumber\\
&&=\chi_{\rm eff.}^{(1)}+\frac{\hbar^2}{2\mu}\left[\frac{\left(\sigma_{0m}-\frac{1}{2}\right)^2-\frac{1}{4}}{r^2}\right],\ \ \mbox{with}\ \ \ A_{+}(\lambda_{ee},\lambda_D)^2\equiv\frac{\pi r_s}{4\Gamma_{ee}}\left(1+\sqrt{1-\frac{24\Gamma_{ee}^2}{\pi r_s(1+\delta^2)}}\right)\ \ \ \mbox{,}\nonumber\\ 
&&A_{-}(\lambda_{ee},\lambda_D)^2\equiv\frac{\pi r_s}{4\Gamma_{ee}}\left(1-\sqrt{1-\frac{24\Gamma_{ee}^2}{\pi r_s(1+\delta^2)}}\right)\ \ \mbox{and}\ \ \sigma_{nm}=n+m+\xi+\frac{1}{2}
\label{EE2}
\end{eqnarray}
\end{widetext}
where we have also incorporate the effect of dynamic screening \cite{KJ1}. $\xi=\phi_{AB}/\phi_{0}$ is taken as integer with the flux quantum $\phi_0=hc/e$. The relative velocity is denoted by $\delta$ and the density parameter is $r_s=\alpha/a_0$ (where $a_0$ is the Bohr radius). 

The purpose of this research is to study entanglement fidelity for low-energy elastic electron-electron interaction in a weak electric field and a strongly coupled semiclassical plasma. Moreover, how can we enhance or generate or simulate a low-energy elastic electron-electron interaction? To answer this question, we need to solve the radial schr\"odinger equation with the effective model (\ref{EE2}) using perturbation theory \cite{SUS1,SUS2}, since the equation does not admit an exact solution with the model. Thus, we obtain the $n$-th state energy as
\begin{widetext}
\begin{eqnarray}
E_{nm}&\approx&-\frac{\mu}{2\hbar^2}\frac{\mathcal{M}^2\left(A_{+}^2-A_{-}^2\right)^2}{\sigma_{nm}^2}-\frac{\hbar^2F}{2\mu\mathcal{M}\left(A_{+}^2-A_{-}^2\right)}\left[3\sigma_{nm}^2-\left(\sigma_{0m}-\frac{1}{2}\right)^2+\frac{1}{4}\right]\nonumber\\
&&\mathcal{M}\left(A_{+}^2A_{-}-A_{+}A_{-}^2\right)+\left\{\left[\frac{\mathcal{M}}{6}\left(A_{+}^2A_{-}^3-A_{+}^3A_{-}^2\right)\right]-\frac{\sigma_{nm}^2\hbar^2F^2}{{2\mu\mathcal{M}^2\left(A_{+}^2-A_{-}^2\right)^2}}\right\}\nonumber\\
&&\times\frac{\sigma_{nm}^2\hbar^4}{{\mu^2\mathcal{M}^2\left(A_{+}^2-A_{-}^2\right)^2}}\left\{5\sigma_{nm}^2-3\left(\sigma_{0m}-\frac{1}{2}\right)^2+\frac{7}{4}\right\}-\frac{\sigma_{nm}^2\hbar^6\left(A_{+}^2A_{-}^4-A_{+}^4A_{-}^2\right)}{96\mu^3\mathcal{M}^2\left(A_{+}^2-A_{-}^2\right)^3}\nonumber\\
&&\times\left[5\sigma_{nm}^2-3\left(\sigma_{0m}-\frac{1}{2}\right)^2+\frac{3}{4}\right]\left[5\sigma_{nm}^2-3\left(\sigma_{0m}-\frac{1}{2}\right)^2+\frac{7}{4}\right]+\frac{\sigma_{nm}^4\hbar^8}{8\mu^4\mathcal{M}^5\left(A_{+}^2-A_{-}^2\right)^5}\nonumber\\
&&\times\left[5\sigma_{nm}^2-3\left(\sigma_{0m}-\frac{1}{2}\right)^2+\frac{7}{4}\right]\left[\frac{\mathcal{M}}{6}\left(A_{+}^2A_{-}^3-A_{+}^3A_{-}^2\right)\right]\left[9\sigma_{nm}^2-5\left(\sigma_{0m}-\frac{1}{2}\right)^2+\frac{5}{4}\right]\nonumber\\
&&-\frac{\sigma_{nm}^6\hbar^{10}}{8\mu^5\mathcal{M}^7\left(A_{+}^2-A_{-}^2\right)^7}\left[9\sigma_{nm}^2-5\left(\sigma_{0m}-\frac{1}{2}\right)^2+\frac{5}{4}\right]\left[5\sigma_{nm}^2-3\left(\sigma_{0m}-\frac{1}{2}\right)^2+\frac{7}{4}\right],
\label{EE29}
\end{eqnarray}
\end{widetext}
where $\mathcal{M}=-{\alpha\Gamma_{ee}}/{\sqrt{1-{24\Gamma_{ee}^2}/({\pi r_s\left(1+\delta^2\right)})}}$. Table 1 displays eigenvalues for electron-electron interaction in semiclassical plasma under the influence of external fields (AB-flux field and electric field) in atomic units and in low vibrational $n$ and rotational $m$ states. When there are no external fields, (i.e., when $\xi = F =0$), the spacing between the energy levels of the effective potential is narrow and decreases as $n$ increases. We found that, there exist degeneracy among some states $(n,m)$, for instance $(0,1)$ and $(2,-1)$; $(1,1)$ and $(3,-1)$. There also exist quasi-degeneracy of the energy levels in $(2,0)$ and $(1,1)$; $(3,0)$ and $(2,1)$.  Not only does the energy levels of the effective potential and spacings between states increases by exposing the system to external fields, but also the degeneracies are removed and become split up and down.
\begin{table}[!t]
{\scriptsize
\caption{\footnotesize The energy values for electron-electron interaction under the influence of AB-flux and external electric fields with various values of magnetic quantum numbers. The following fitting parameters have been employed: $\alpha=r_s=5$, $\Gamma_{ee}=1$,  $\delta=8$ and $\hbar=\mu=c=1$  .} \vspace*{10pt}{
\begin{tabular}{ccccccc}\hline\hline
{}&{}&{}&{}&{}&{}&{}\\[-1.0ex]
m	&n	&$F=0$, $\xi=0$&$F=0$, $\xi=2$,	&$F=0.01$, $\xi=0$	&$F=0.01$, $\xi=2$\\[1ex]\hline
0	&0	&-3092.11        &-131.152        &-3092.11   & -131.150\\[1ex]
		&1	&-350.535       &-70.4042     &-350.534  &-70.4000\\[1ex]
		&2	&-131.104         &-44.5892     &-131.102    &-44.5819\\[1ex]
		&3	&-70.2885       &-29.7294     &-70.2838  &-29.7183\\[1ex]
{}	&{}&		{}				       &    {}            &{}              &            {}\\[1ex]						
1		&0	&-350.538	     &-70.5367      &-350.538  &-70.5331\\[1ex]
		&1	& -131.117         &-44.8667     & -131.114    &-44.8601\\[1ex]
		&2	&-70.3183        &-30.2518      & -70.3137  &-30.2413\\[1ex]
		&3	&-44.4121        &-18.6248      &-44.4044  &-18.6098\\[1ex]
{}	&{}&		{}				       &    {}            &{}              &            {}\\[1ex]						
-1	&0	& -3092.11       &-350.538	     & -3092.11	  &-350.538\\[1ex]
		&1	& -3092.11       &-131.117         &-3092.11    &-131.114\\[1ex]
		&2	&-350.538	     &-70.3183       &-350.538  &-70.3137\\[1ex]
		&3	&-131.117	         &-44.4121       &-131.114	    &-44.4044\\[1ex]
	\hline\hline
\end{tabular}\label{tab2}}
\vspace*{-1pt}}
\end{table}
It has been shown that increasing the strenght of AB-flux field would lead to a substantial shift in the energies. Its important to note that the dominance and confinement effects of   AB-flux field  on electron-electron interaction in semiclassical plasma is stronger than the electric field and as consequence the localizations of the quantum states can be manipulated via application of AB-flux field

It can also be seen from this table that the lowest energy can be obtained when $n = 0|1$, $m= -1$, $\xi=0$ and $F=0|0.01$. However we can do a more thorough analysis. Suppose $\xi=0.4$ and $F=1$, we have  $E_{0,-1}=-77114.1$a.u. For $\xi=2$ and $F=1$, we have $E_{0,-1}=-350.462$a.u.  Such a large discrepancy in the energies reveal that the effect of AB flux field is more dominance on the system than the electric field. Consequently, we focus on $\xi$ and we found that for $\xi\rightarrow0.5$, for any $F$, $E_{0,-1}\rightarrow-\infty$.  In addition, for $\xi=0.7$ and $F=1$, we have $E_{0,-1}=-19284.4$a.u. This indicates that the lowest $E_{nm}$ would be obtained when $m=-1$ and $\xi \rightarrow0.5$. 

Now, to study the entanglement fidelity in the interaction process under the influence of external electric field and in a strongly coupled semiclassical plasmas, let's define the entanglement  fidelity (which is the overlap between maximally entangled state) integrated over the internal coordinates $\vec{R}$ and $\vec{r}$ as \cite{JA11} $f(\tau)=\left|\left.\left\langle \psi_{me}(\vec{R},\vec{r})\right.\right.\right|\left.\left.\psi_{ee}(\vec{R},\vec{r},\tau)\right\rangle\right|^2$ where 
\begin{eqnarray}
\psi_{ee}(\vec{R},\vec{r},\tau)&=&(2\pi)^{-3/2}\exp\left[-i\left(\frac{P^2}{2M}+E\right)\tau\right]\nonumber\\
&&\times\exp(i\vec{P}\cdot\vec{R})\Psi(\vec{r})
\end{eqnarray}
 denotes the wavefunction of the bipartite interaction process, $\vec{P}$ is the momentum of the electron-electron interaction and 
\begin{equation}
\psi_{me}(\vec{R},\vec{r})=(2\pi)^{-3/2}\int d^3k\exp\left[i\vec{k}\cdot\vec{r}_1\right]\exp(i\vec{k}\cdot\vec{r}_2).
\label{EE30}
\end{equation}
With these expressions, the entanglement fidelity can be written as square of the scattered wave function for a given model; i.e., $f_{\vec{k}}\propto\left|d^3{\bf r}\Psi_{\vec{k}}({\bf r})\right|^2$ \cite{JA11}, where $\Psi_{\vec{k}}({\bf r})$ is the solution of the scattered wave function which can be expressed in terms of partial wave expansion as
\begin{equation}
\Psi_{\vec{k}}({\bf r})=(2\pi)^{-3/2}\sum_{m+\xi=\frac{1}{2}}^\infty(2m+2\xi)i^{g}\mathcal{T}^k_g R_g^k(r)P_g(cos\theta),
\label{EE31}
\end{equation}
where the wave number is $k=\sqrt{2\mu E/\hbar^2}$, $g=m+\xi+1/2$, $\mathcal{T}^k_g$ represents the expansion coefficient, $R_g^k(r)$ denotes the solution of the radial wave equation and the Legendre polynomial is denoted as $P_g(cos\theta)$. For a spherical symmetric potential ($\chi_{\rm eff.}^{(1)}$), the following expressions 
\begin{eqnarray}
&&\mathcal{T}^k_g=\left[1+\frac{2i\mu k}{\hbar^2}\int_0^\infty j_g(kr)\chi_{\rm eff.}^{(1)}R_g^k(r)r^2dr\right]^{-1}\ \ \ \mbox{and}\nonumber\\
&&\left[\frac{d^2}{dr^2}+\frac{2}{r}\frac{d}{dr}+k^2-\frac{2\mu}{\hbar^2}\left(\chi_{\rm eff.}^{(1)}+\frac{\hbar^2}{2\mu}\left[\frac{g(g-1)}{r^2}\right]\right)\right]R_g^k(r)\nonumber\\
&&=0,
\label{EE32}
\end{eqnarray} 
hold for the expansion coefficient and the radial wave equation respectively. The eigenfunction $R_g^k(r)$ can be expressed in terms of spherical Neumann function and spherical Bessel function as follows:
\begin{eqnarray}
R_g^k(r)&=&j_g(kr)+\frac{2\pi k}{\hbar^2}\left[n_g(kr)\int_0^r dr r^2j_g(kr)\chi_{\rm eff.}^{(1)}R_g^k(r)\right.\nonumber\\
&&\left.+j_g(r)\int_r^\infty dr r^2n_g(kr)\chi_{\rm eff.}^{(1)}R_g^g(r)\right].
\label{EE33}
\end{eqnarray} 
As earlier stated, the lowest energy can be obtained by setting $m=-1$ and $\xi\approx0.5$, hence, we write the entanglement fidelity for low-energy elastic interaction by using equations (\ref{EE32}) and (\ref{EE33}) to find
\begin{eqnarray}
f_{\vec{k}}&\propto&\frac{\left|\int_0^\infty d{r}r^2R_0^k(r)\right|^2}{1+\left|\frac{2\mu k}{\hbar^2}\int_0^\infty d{r}r^2\chi_{\rm eff.}^{(1)}R_0^k(r)\right|^2}\nonumber\\
&\propto& \frac{\left|\int_0^\infty d{r}r^2\frac{\sin(kr)}{kr}\right|^2}{1+\left|\frac{2\mu k}{\hbar^2}\int_0^\infty d{r}r^2\chi_{\rm eff.}^{(1)}\frac{\sin(kr)}{kr}\right|^2}.
\label{EE34}
\end{eqnarray} 
Thus, we express the entanglement fidelity for electron-electron interaction under the influence of electric field and surrounded by strongly coupled semiclassical plasmas, in terms of the ratio of entanglement fidelity for model \ref{EE2} to that of pure Coulomb potential as:
\begin{widetext}
\begin{eqnarray}
R_K^{(\lambda_{ee},\lambda_D)}&=&\frac{1+\left|-\frac{2\mu Z^2e^2k}{\hbar^2}\int_0^\infty dr r \frac{\sin(kr)}{kr}\right|^2}{1+\left|{-\frac{2\mu\mathcal{M} k}{\hbar^2}\int_0^\infty dr r^2\left[\left(A_{-}^2\frac{e^{-Ar}}{r}-A_{+}^2\frac{e^{-Br}}{r}\right)-Fr+\frac{\omega_c\hbar}{2}\left(\sigma_{0m}-\frac{1}{2}\right)+\left(\frac{\mu\omega_c^2}{8}\right)r^2\right]\frac{\sin(kr)}{kr}}\right|^2}\nonumber\\
&=&\left(\epsilon+1\right)\left({\epsilon+\epsilon^2\tilde{\mathcal{M}}^2\left[\frac{A_{-}^2}{A_{+}^2+k^2}-\frac{A_{+}^2}{A_{-}^2+k^2}-\frac{2F}{k^4}\right]^2}\right)^{-1},
\label{EE35}
\end{eqnarray}
\end{widetext}
where $\epsilon$ is the projectile energy and it is given by $\epsilon=E/(13.6Z_e^4)$ and $\tilde{\mathcal{M}}=2\mu^2z^2e^2\mathcal{M}/\hbar^4$. In figure \ref{fig1}, we have plotted the variation of $R_K^{(\lambda_{ee},\lambda_D)}$ as a function of projectile energy. In (a) we found that for a weak electric field, the fidelity ratio for electron-electron interaction increase as projectile energy increase but remains constant or almost zero for a strong electric field.  That is to say, a weak electric field intensifies entanglement fidelity in a strong coupled semiclassical plasmas. We use the result of (a) to proceed to (b) (i.e. we fixed $F$ at $0.02$) by studying the fidelity ratio as a function of projectile energy for various values of $\alpha$ which is average distance between the particles.  As it can be seen, increasing the distance between the electrons has infinitesimal effect on the fidelity ratio. Moreover distorting the intensity electric field (say, $F=0.02\rightarrow0.2$) will results into a large discrepancy. These results provide us a valuable information on how the efficiency of entanglement fidelity for a low-energy elastic electron-electron interaction in a strong semiclassical plasma can be influenced by the presence of external fields.
\section{Concluding Remarks} 
We found that to maintain a low-energy elastic electron-electron interaction in strongly semiclassical plasma, weak external fields are prerequisite. The entanglement fidelity in the interaction process has been explored. We have used partial wave analysis to derive the entanglement fidelity. We found that for a low electric field, the entanglement fidelity varies with projectile energy. Our results provide a valuable information on how the efficiency of entanglement fidelity for a low-energy elastic electron-electron interaction in a strong semiclassical plasma can be influenced by the presence of external field.


\end{document}